\documentclass[prl,twocolumn,showpacs,floatfix,amsmath,amssymb,superscriptaddress]{revtex4-1}
\usepackage{amsfonts}
\usepackage{stmaryrd}
\usepackage{bbm}
\usepackage{mathrsfs}
\usepackage{tipa}
\usepackage{amssymb}
\usepackage{txfonts}
\usepackage{graphicx}
\usepackage{dcolumn}
\usepackage{latexsym}
\usepackage{epstopdf}
\usepackage{amsmath}
\usepackage[colorlinks,linkcolor=blue,urlcolor=blue,citecolor=blue]{hyperref}
\usepackage{multirow}
\usepackage{subfigure}
\usepackage{url}
\newcommand{\cm}{cm$^{-1}$}

\begin{document}
\bibliographystyle{apsrev4-1}

\title{Band structure reconstruction in the
topological semimetal PrAlSi} 

\author{B. X. Gao}
\affiliation{Center for Advanced Quantum Studies and Department of Physics, Beijing Normal University, Beijing 100875, China}
\affiliation{Key Laboratory of Multiscale Spin Physics, Ministry of Education, Beijing Normal University, Beijing 100875, China}

\author{M. Lyu }
\affiliation{Beijing National Laboratory for Condensed Matter Physics, Institute of Physics, Chinese Academy of Sciences, Beijing 100190, China}

\author{L. Y. Cao}
\affiliation{Center for Advanced Quantum Studies and Department of Physics, Beijing Normal University, Beijing 100875, China}
\affiliation{Key Laboratory of Multiscale Spin Physics, Ministry of Education, Beijing Normal University, Beijing 100875, China}

\author{L. Wang}
\affiliation{Center for Advanced Quantum Studies and Department of Physics, Beijing Normal University, Beijing 100875, China}
\affiliation{Key Laboratory of Multiscale Spin Physics, Ministry of Education, Beijing Normal University, Beijing 100875, China}

\author{X. T. Zhang}
\affiliation{Center for Advanced Quantum Studies and Department of Physics, Beijing Normal University, Beijing 100875, China}
\affiliation{Key Laboratory of Multiscale Spin Physics, Ministry of Education, Beijing Normal University, Beijing 100875, China}

\author{X. Y. Zhang}
\affiliation{Center for Advanced Quantum Studies and Department of Physics, Beijing Normal University, Beijing 100875, China}
\affiliation{Key Laboratory of Multiscale Spin Physics, Ministry of Education, Beijing Normal University, Beijing 100875, China}

\author{P. J. Sun}
\affiliation{Beijing National Laboratory for Condensed Matter Physics, Institute of Physics, Chinese Academy of Sciences, Beijing 100190, China}
\affiliation{School of Physical Sciences, University of Chinese Academy of Sciences, Beijing 100049, China}
\affiliation{Songshan Lake Materials Laboratory, Dongguan, Guangdong 523808, China}

\author{R. Y. Chen}
\affiliation{Center for Advanced Quantum Studies and Department of Physics, Beijing Normal University, Beijing 100875, China}
\affiliation{Key Laboratory of Multiscale Spin Physics, Ministry of Education, Beijing Normal University, Beijing 100875, China}

\begin{abstract}
The interplay between nontrivial topology, magnetism and strong correlation has generated considerable research interest in condensed matter physics. The topological RAlX (R = rare earth ; X = Si and Ge) family has provided an excellent platform for exploring these complex interactions. Here, we performed infrared spectroscopy measurements on the  ferromagnetic (FM) topological semimetal PrAlSi, in oder to investigate the impact of FM orderings on the topological band structure. We find that the optical conductivity associated with the Dirac/Weyl cones exhibits two segments of linearly increasing parts in the normal state, connected by a kink feature at around 1 960  \cm. By entering the FM state, however, an additional linear-growing segment shows up in between the original ones, suggesting that the band structure is reconstructed. We propose that these observations can be effectively explained by a scenario where the Dirac/Weyl nodes are split into pairs of Weyl nodes with lower degeneracy, due to the time reversal symmetry breaking induced by the FM ordering. This band structure reconstruction also leads to a sudden enhancement of the itinerant carrier density. In addition, the effective mass of the itinerant carriers are estimated to be two orders of magnitude smaller than the free electron mass, providing a rare case where nearly all the free carriers exhibit behaviors characteristic of relativistic Dirac or Weyl fermions. Our results demonstrate an compelling example of the strong interaction between magnetic order and topological band structures, which opens up new avenues for exploring novel topological materials and their potential applications.   
\end{abstract}

\maketitle

\section{Introduction}
The discovery of relativistic quasiparticles such as Dirac and Weyl fermions in topological materials is one of the most important achievements in condensed matter physics in the last two decades \cite{Novoselov_Two-dimensional_2005,Castro_The_2009,Wan_Topological_2011,Su_Discovery_2015,Lv_Experimental_2015}. The band structure of Dirac fermions exhibits linear dispersion and crosses at the so-called Dirac point. It is well known that the four-fold degenerate Dirac point is protected by both space inversion symmetry (SIS) and time reversal symmetry (TRS). When space inversion symmetry is broken, as in noncentrosymmetryic materials, the Dirac point will be split into two Weyl points which locate at different energies. When time reversal symmetry is broken, either by external magnetic field or internal magnetic orderings, the Dirac point will turn into a pair of Weyl points that separate in the momentum space. Particularly, Weyl semimetals could be further categorized into type-I and type-II depending on the tilting of the Weyl cones \cite{Carbotte_Dirac_2016}. 
In reality, noncentrosymmetric (such as TaAs \cite{Su_Discovery_2015,Lv_Experimental_2015}, TaP \cite{Xu_Experimental_2015,li_concurrence_2017}, and NbP \cite{Shekhar_Extremely_2015,Anna_Chiral_2016} etc.) and magnetic (such as YbMnBi$_2$ \cite{Wang_Magnetotransport_2016,Borisenko_Time-reversal_2019}, and Mn$_3$Sn \cite{Kuroda_Evidence_2016,Tsai_Electrical_2020} etc.) Weyl semimetals are both intensively investigated. However, materials with simultaneous broken SIS and TRS are relatively rare.


Recently, the rare earth-based compounds RAlX have emerged as an ideal system to investigate Weyl-related physics, featuring broken SIS and/or TRS, different types of Weyl fermions, and even Kondo effect.
The lattice structure of RAlX was initially identified to belong to the noncentrosymmetric $I4_1md$ ($C_{4v}$) space group \cite{Guloy_Syntheses_1991}, which lacks space inversion symmetry, but subsequent studies discover that the mixing between Al and X site can turn the compound into centrosymetric $I4_1amd$ ($C_{4h}$) group \cite{lyu_Nonsaturating_2020}, where the SIS is preserved. As a result, both centrosymetric and noncentrosymetric lattice structures are reported in some of the RAlX samples, such as PrAlSi \cite{Svilen_Ternary_2005,A.L._Oscillations_2007, lyu_Nonsaturating_2020,Yang_Transition_2020}, and CeAlSi \cite{Svilen_Ternary_2005,A.L._Oscillations_2007,Yang_Noncollinear_2021}. Moreover, various magnetic structures have been observed in the RAlX compounds, except for LaAlSi and LaAlGe. 
For instance, CeAlSi and PrAlSi experience FM phase transitions \cite{Yang_Noncollinear_2021,lyu_Nonsaturating_2020}, while antiferromagnetic (AFM) order is developed in GdAlSi 
 \cite{Laha_Electronic_2024}. 
Particularly, NdAlGe \cite{Zhao_Field-induced_2022}, NdAlSi \cite{Gaudet_Weyl-mediated_2021,Wang_Ndalsi_2022} and SmAlSi \cite{Lou_Signature_2023,Yao_Large_2023} exhibit more complex configurations with multiple magnetic phases. TRS breaking are thus enabled in those with FM orderings and intertwined with SIS breaking in these materials. As a results, special Weyl fermions are realized in these materials, and both type I and type II Weyl fermions are reported to exist in LaAlSi \cite{Su_Multiple_2021}, PrAlGe \cite{Chang_Magnetic_2018}, and CeAlGe \cite{Chang_Magnetic_2018}.
Furthermore, due to the existence of 4$f$ electrons in the rear-earth element, Kondo effect is proposed to be prominent in CeAlGe, leading to the realization of the so-called Weyl-Kondo system \cite{Corasaniti_Evidence_2021}.  Besides, a series of anomalous responses are demonstrated in the RAlX family, including topological Hall effect (THE) \cite{Puphal_Topological_2020,Yao_Large_2023}, anomalous Hall effect (AHE) \cite{Meng_Large_2019,lyu_Nonsaturating_2020,Sanchez_Observation_2020,Destraz_Magnetism_2020,Yang_Noncollinear_2021,Piva_Topological_2023,Laha_Electronic_2024}, anomalous Nernst effect (ANE) \cite{Destraz_Magnetism_2020}, loop Hall effect (LHE) \cite{Yang_Noncollinear_2021,Piva_Topological_2023}, etc.

As a member of the magnetic RAlX family, PrAlSi experiences a FM transition at $T_{C}$ = 17.8 K  \cite{lyu_Nonsaturating_2020}, and two weak reentrant magnetic phase transitions at $T_{M1}$ $\simeq$ 16.5 K and $T_{M2}$ $\simeq$ 9 K are revealed by M. Lyu et al., which are proposed to be spin glasses or ferromagnetic cluster glasses  \cite{lyu_Nonsaturating_2020}. Huge AHE \cite{lyu_Nonsaturating_2020} and anisotropic magnetocaloric effect (MCE) \cite{Lyu_Large_2020} are observed in PrAlSi, associated with its nontrivial band structure. Moreover, magnetotransport measurements reveal a rare field-induced Lifshitz transition (LT) at 14.5 T, which involves the hole-like Weyl pockets along the $\Gamma$-X direction \cite{Wu_Field-induced_2023}. Across the FM transition, the Shubnikof-de Haas (SdH) oscillation results indicate that the Fermi surface (FS) of PrAlSi increases by nearly 40\% when the temperature drops from slightly above $T_{C}$ to 2 K \cite{lyu_Nonsaturating_2020}. However, the angle resolved photo-emission spectroscopy measurements together with first-principles calculations demonstrate that the low-energy band structure of PrAlSi has almost no obvious evolution across $T_C$, suggesting the coupling between the 4$f$ electrons and the conduction electrons is probably negligible  \cite{Lou_Signature_2023}. 
In order to further reveal the interplay between the itinerant carriers and the local 4$f$ moments in PrAlSi, we systematically studied the charge dynamics of this compounds using infrared spectroscopy. It is found that the band structure of PrAlSi is strongly modified across the FM phase transitions, which not only turns the Dirac/Weyl nodes into pairs of Weyl nodes with lower degeneracy, but also gives rise to a sudden enhancement of the carrier density. 

\section{Experimental}

Single crystals of PrAlSi were grown by self-flux in the high-temperature molten Al solvent  \cite{lyu_Nonsaturating_2020}.  The optical reflectivity measurement was performed on a Bruker Vertex 80 V spectrometer in the frequency range 40 – 25 000 \cm. In order to obtain the exact reflectivity \emph{R($\omega$)} value, we use a technology of in-situ coating of a gold or aluminum film on the sample. The real part of the optical conductivity $\sigma_{1}$($\omega$) is obtained by the Kramers-Kronig transform of \emph{R($\omega$)}. We use the Hagen-Rubens relationship to extrapolate in the low-frequency part of \emph{R($\omega$)}, and  we use the x-ray atomic diffraction function to extrapolate in the high-frequency part  \cite{Tanner_Use_2015}.

\section{Results and Discussions}



\begin{figure}[tb]
    \centering
    \includegraphics[scale=0.95]{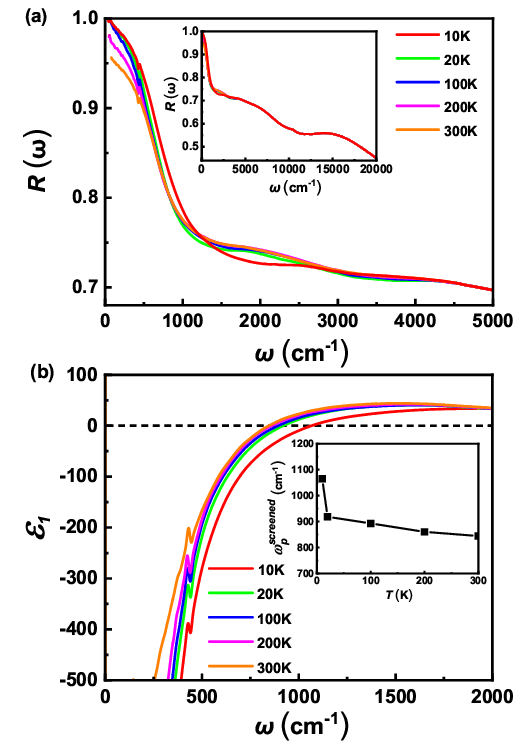}
    \caption{(a) The temperature-dependent optical reflectivity \emph{R($\omega$)} for PrAlSi below 5 000 \cm. The inset displays \emph{R($\omega$)} in a large energy scale up to 20 000 \cm. (b) The temperature-dependent dielectric permittivity $\varepsilon_{1}(\omega)$. The inset shows the temperature dependence of the screened plasma frequency $\omega _{p}^{screened}$.}
    \label{fig1}
\end{figure}

The frequency-dependent reflectivity \emph{R($\omega$)} of PrAlSi at several selected temperatures is shown in Fig.\ref{fig1}(a). In the far-infrared region where the frequency gets close to zero, \emph{R($\omega$)} approaches unity and increases as temperature decreases, which are typical characteristics of metallic materials. With frequency increasing, \emph{R($\omega$)} exhibits a steep drop  and reaches a dip-like feature near 1 500 \cm, which is another hallmark of good metal, i.e. the so-called plasma edge. The position of the plasma edge corresponds to the screened plasma frequency $\omega _{p}^{screened}$  \cite{Degiorgi_Low-temperature_1997}, which is associated with the plasma frequency $\omega_{p}$ by $\omega _{p}^{screened} =\omega_{p}/\sqrt{\varepsilon_{\infty } }$, with $\sqrt{\varepsilon_{\infty } } $ 
being the dielectric constant at high energy. Notably, the plasma frequency is related to the carrier density $n$ and effective mass $m^*$ by $\omega_{p} =\sqrt{4\pi ne^{2}/m^{*}}$. Therefore, the evolution of the plasma edge can provide essential information on the itinerant quasiparticles. It can be seen in Fig.\ref{fig1}(a) that the position of the plasma edge is quite stable against temperature variation above $T_C$. Meanwhile, it gets slightly sharper with temperature decreasing, which indicates a continuous reducing scattering rate. Across the FM phase transition, however, the plasma edge shows a sudden jump to higher energy, which suggests that some properties of the free carriers are strongly modified.

\begin{figure*}[btp]
    \centering
    \includegraphics[scale=0.72]{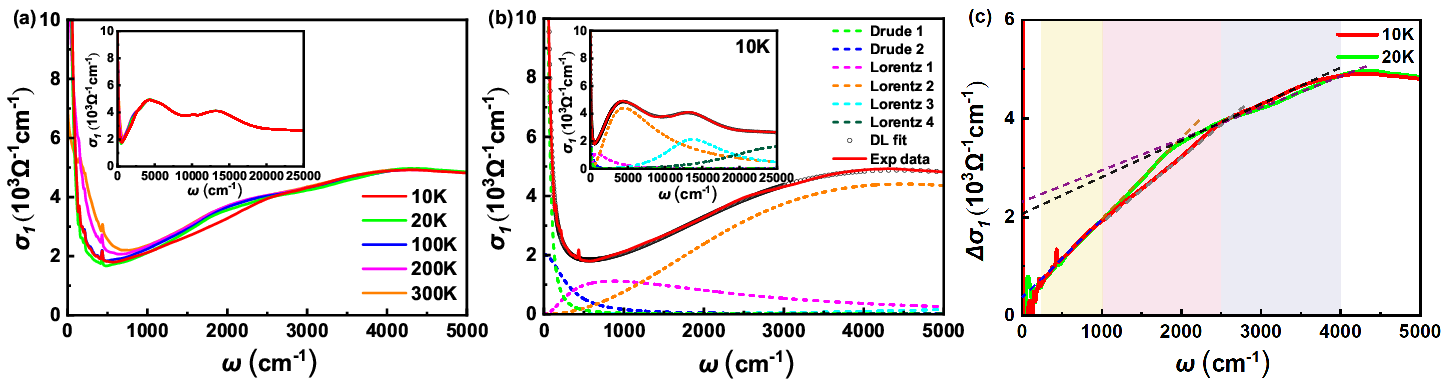}
    \caption{(a) The temperature-dependent  optical conductivity $\sigma_{1}(\omega$) for PrAlSi below 5 000 \cm. The inset displays $\sigma_{1}(\omega$) in a large energy range up to 20 000 \cm. (b) The Drude-Lorentz fitting results of $\sigma_{1}$($\omega$) at 10 K below 5 000 \cm. The inset displays the the fitting results up to 20 000 \cm. (c) The
    interband transition contributed optical conductivity $\Delta\sigma_{1}$($\omega$) at 10 K and 20 K, shown as red and green solid lines, respectively. The dashed lines indicate the linear frequency dependence of $\Delta\sigma_{1}$($\omega$) in different energy intervals. The three linearly increasing segments at 10 K are highlighted by the pale-yellow, pale-pink and pale-purple regions.}
    \label{fig2}
\end{figure*}

In order to further clarify the abrupt change of the plasma edge, we also obtain the frequency dependent dielectric permittivity $\varepsilon_{1}(\omega)$, as shown in Fig.\ref{fig1}(b). It is well known that $\varepsilon_{1}(\omega)$ of a metallic compound crosses zero at the screened plasma frequency  $\omega _{p}^{screened}$  \cite{Martin_Electrodynamics_2002}. By this method, the exact values of $\omega _{p}^{screened}$ could be obtained and is plotted in the inset of Fig.\ref{fig1}(b). It is clearly shown that $\omega _{p}^{screened}$ increased very slightly upon cooling until $T_C$, below which a significant enhancement is observed, in line with the sudden blue-shift of the plasma edge. The enhancement of  $\omega _{p}^{screened}$ suggests either an abrupt enhancement of the carrier density $n$ or decrease of the effective mass $m^{*}$. As the effective mass is related to the correlation strength, which is usually robust against temperature changing and magnetic transitions, it is most likely that the carrier density is enhanced across $T_C$. This conclusion is consistent with the quantum oscillation experiments which demonstrate an enlarged FS in the FM state \cite{lyu_Nonsaturating_2020}.


The evolution of overall physical properties of PrAlSi can be characterized more clearly through the optical conductivity. Fig.\ref{fig2}(a) shows the real part of the optical conductivity $\sigma_{1}(\omega$) below 5 000 \cm at different temperatures in its main panel, while the inset presents $\sigma_{1}(\omega$) in the energy scale up to 20 000 \cm. Pronounced Drude responses are observed in the whole temperature range, i.e. a peak centered at zero frequency, which is contributed by excitation of the itinerant carriers. The half width of the Drude peak represents the scattering rate while the spectral weight (SW) of it is related to the plasma frequency $\omega_P$. As $T$ decreases, the Drude peak narrows, reflecting the decrease of the quasiparticle scattering rate, which agrees with the sharpening of the plasma edge in the reflectivity spectra. Meanwhile, the variation of SW is not very straightforward, which will be discussed later. 

Above the Drude response energy, $\sigma_{1}(\omega)$ shows quasi-linear increasing behaviors as a function of frequency $\omega$. Linear increasing $\sigma_{1}(\omega)$ is usually considered as as hallmark of the existence of three-dimensional (3D) Dirac or Weyl fermions. In order to test the credibility of this property, it is necessary to remove the Drude responses from the total excitation. To achieve this goal, we first use the Drude-Lorentz model to decompose $\sigma_{1}$($\omega$):
\begin{equation}
    \varepsilon _{\omega } =\varepsilon_{\infty}-\sum_{s}^{}\frac{\omega _{ps}^{2} }{\omega^{2}-i\omega/\tau _{Ds}  }+\sum_{j}^{} \frac{S_{j}^{2} }{\omega_{j}^{2} -\omega^{2}-i\omega/\tau _{j}}
\label{equ-01}
\end{equation}
here, $\varepsilon_{\infty}$ is the dielectric constant at high energy; the middle term is the Drude component that characterizes the electrodynamics of itinerate carriers, and the last term is the Lorentz component which describes excitations across energy gaps or interband transitions. The experimental $\sigma_{1}(\omega$) can be well reproduced by two Drude terms and several Lorentz terms, as shown in Fig.\ref{fig2}(b). Thus, the interband transition contributions can be yielded by subtracting the Drude term from the experimental data (i.e., $\Delta\sigma_{1}$($\omega$) = $\sigma_{1}$($\omega$) - $\sigma_{1}^{Drude}$($\omega$)). The obtained $\Delta\sigma_{1}$($\omega$) at 10 K and 20 K are displayed in Fig.\ref{fig2}(c), by red and green solid lines, respectively.  Linearly increasing segments are clearly observed both above and below $
T_C$. However, notable discrepancies arise between these two temperatures. $\Delta\sigma_{1}$($\omega$) at 20 K is a representative of the normal state situation, which demonstrates perfectly linear relationship with frequency $\omega$ in two separate segments, connected by a kink at around 1 960 \cm. By contrast, $\Delta\sigma_{1}$($\omega$) at 10 K exhibit three linearly increasing parts, as highlighted by the shaded areas of different colors. Moreover, the first and third linear parts of 10 K almost overlap with the two linear segments of 20 K, as can be seen in the pale-yellow and pale-purple area in Fig.\ref{fig2}(c). Meanwhile, the additional linearly-increasing part appearing only at 10 K stands out in the pale-pink background, and two kinks are observed at 1 015 \cm and 2 490 \cm, respectively. This significant change in optical conductivity strongly indicates a band structure reconstruction across the FM transition in PrAlSi.

Now that we have confirmed the existence of linearly growing optical conductivity in a large energy range, 3D Dirac or Weyl fermions are expected to dominate the charge dynamics of PrAlSi around the Fermi level. It is worth to remark that both noncentrosymmetric LaPtSi type and centrocenmetryic $\alpha$-ThSi stype structures are reported for PrAlSi \cite{Svilen_Ternary_2005,A.L._Oscillations_2007, lyu_Nonsaturating_2020,Yang_Transition_2020}. Therefore, it is difficult for us the determine whether PrAlSi is a Dirac or Weyl semimetal in the normal state.
Notably, our results of $\Delta\sigma_{1}$($\omega$) at 20 K agrees well with a recent infrared study of PrAlSi performed at room temperature \cite{Kunze_Optical_2024}, which adopts the noncentrosymmetric Weyl scenario. Moreover, as the two segments of linear increasing optical conductivity are connected by a kink and the higher energy segment shows a smaller slope, they are considered as evidences of type-II Weyl fermions. Although this interpretation can adequately explain our results in the normal state, the appearance of an additional linearly growing $\Delta\sigma_{1}$($\omega$) in the FM state remains unclear.  

\begin{figure}[h]
    \centering
    \includegraphics[scale=0.55]{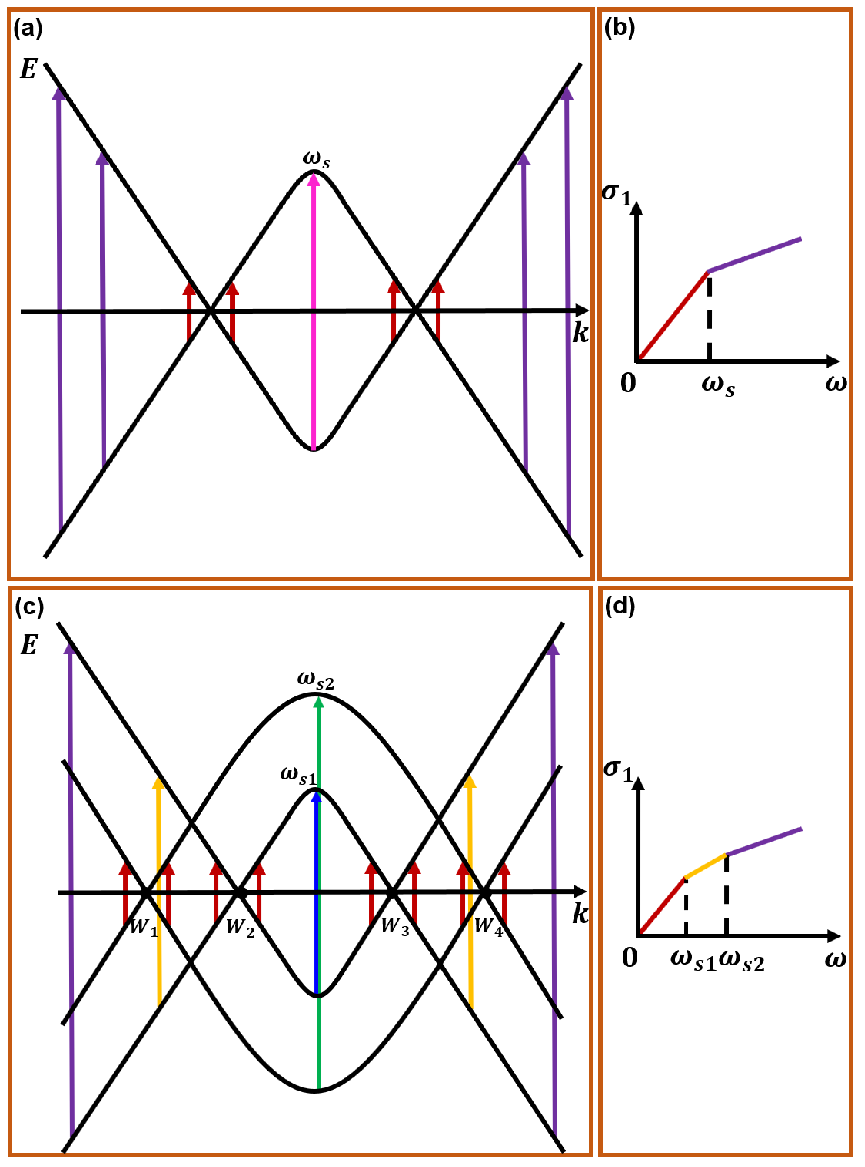}
    \caption{Schematic model of the band structure reconstruction. Panel (a) illustrate the interband transitions linked to two adjacent Dirac/Weyl cones, while the corresponding optical conductivity is displayed in panel (b). The red and purple arrows in panel (a) indicate the transition occuring at energies below and above $\omega_S$, which is marked by the pink arrow. The respective optical conductivity are represented by red and purple solid lines in panel (b). Panel (c) displays the interband transitions in the FM state, where the two adjacent Dirac/Weyl nodes shown in panel (a) are split into four Weyl nodes, labeled as $W_{1}$, $W_{2}$, $W_{3}$, and $W_{4}$. The blue and green solid arrows denote the two singularity points $\omega_{s1}$ and $\omega_{s1}$, separately. The red and purple arrows indicates transitions occurring below $\omega_{s1}$ and above $\omega_{s2}$ respectively, while the yellow arrows represent transitions ocurring between $\omega_{s1}$ and $\omega_{s2}$. The corresponding optical conductivity is depicted in panel (d), where the linearly increasing solid lines are colored to match the transition arrows in panel (c).}
    \label{fig3}
\end{figure}

Alternatively, there is another approach which can account for our data regardless of the exact type of the relativistic quarsiparticles. We notice that there are multiple pairs of Weyl points predicted by theoretical calculation of PrAlSi \cite{Yang_Transition_2020}. Especially, some of the Weyl points are close to each other in the momentum space. Although this calculation is based on the noncentrosymmetric structure, which would split the Weyl nodes into different energy levels, it is not unreasonable to assume that there would be multiple adjacent Dirac points in the band structure of PrAlSi with the centrosymmetric structure. In this case, the neighbouring Dirac or Weyl fermions would contribute to the optical conductivity in a cooperative way.  C. J. Tabert and J. P. Carbotte have calculated the optical conductivity of a similar band structure, and found that the optical conductivity would show two segments of linearly increasing parts connected by a kink at $\omega'$ and it rises much more slowly above this energy \cite{Tabert_Optical-conductivity_2016}.
As illustrated in Fig.\ref{fig3}(a), when there are adjacent Dirac/Weyl nodes around the Fermi level, interband transitions associated with these Dirac/Weyl cones could take place at various momentum positions as illustrated by the red solid arrows in Fig.\ref{fig3}(a), which give rise to an overall linear optical conductivity at low energies as demonstrated by the red segment in Fig.\ref{fig3}(b). However, the transitions between the two neighboring Dirac/Weyl points would reach a maximum, as denoted by the pink solid arrow in Fig.\ref{fig3}(a), which can be treated as a van-hove singularity. Above this singularity energy $\omega_{s}$, the interband transitions are only available on the outer side of the Dirac/Weyl nodes, as represented by the purple solid arrows in Fig.\ref{fig3}(a), whereas the area between the two adjacent Dirac/Weyl nodes are prohibited. As a result, the slope of the linearly increasing optical conductivity is reduced at higher energies, as sketched by the purple segment in  Fig.\ref{fig3}(b). The overall optical conductivity agrees perfectly with our results of PrAlSi in the normal state.

The charge dynamics in the FM state could be explained by this scenario as well. Since the TRS is broken by the FM order, the Dirac/Weyl nodes will further split into Weyl nodes with lower degeneracy, which are marked as $W_{1}$, $W_{2}$, $W_{3}$, and $W_{4}$, as depicted in Fig.\ref{fig3}(c). In this case, there will be two singularity points at the transition energy of $\omega_{s1}$ and $\omega_{s2}$, as denoted by blue and green solid arrows. The corresponding optical conductivity contributed by interband transitions are sketched in Fig.\ref{fig3}(d). Below $\omega_{s1}$, interband transitions are allowed in all of the four split Weyl cones, as indicated by the red solid arrows, therefore the optical response coincides with that of the normal state. Between $\omega_{s1}$ and $\omega_{s2}$, no transitions could take place in the area between $W_2$ and $W_3$, therefore the slope of linearly increasing optical conductivity reduces slightly, as represented by the yellow segment in Fig.\ref{fig3}(d). Above $\omega_{s2}$, the inner part between $W_1$ and $W_4$ are further forbidden, and the optical response overlaps with the normal state once again. These features are in excellent consistence with our results in the FM state. Therefore, we believe the band structure of PrAlSi is reconstructed across the FM phase transition, and the the Dirac/Weyl nodes are split into pairs of Weyl nodes with lower degeneracy. 

It is worth noting that the model illustrated in Fig.\ref{fig3} is only a schematic representation and it is highly simplified. The Dirac/Weyl points are crudely situated exactly at the Fermi level for the sake of simplicity, and factors such as spin-orbital coupling are not taken into account. Moreover, the FM order might not only split the degeneracy of the Dirac/Weyl node, but also induce Zeeman splitting of the electronic bands, which is not considered either. Consequently, the change of itinerant carriers across the FM transition can not be accurately captured.  In reality, 
it is very likely that the band structure reconstruction across $T_C$ also leads to the shift of the Weyl cones relative to the Fermi level, and therefore the expansion or shrinkage of FS(s). Based on our results, the carrier density is suddenly enhanced by entering the FM state, implying the enlargement or emergence of FS(s). However,
the detail band structure of PrAlSi in the FM state is expected to be extremely complex, which is beyond our speculation and highly requires for future theoretical investigations.  

Since some of the RAlGe compounds are proposed to be Kondo-Weyl semimetals \cite{Corasaniti_Evidence_2021}, which means the Kondo coupling between $f$ electrons and the linear dispersive bands strongly modifies the band structures around $E_F$, we would like to discuss weather PrAlSi belong to this category. In order to do so, it is crucial to identify the correlation strength in this material, which can be reflected by several different parameters. Usually, Kondo effect could lead to a downward bending of resistivity, which is unfortunately not observed in PrAlSi \cite{lyu_Nonsaturating_2020}.  Other features of a strong correlation effect include the reduction of Fermi velocity $V_{F}$ and the enhancement of the effective mass $m^{*}$, which will be discussed below.

It is proposed that 
the low-temperature renormalized $V_{F}$ of two- and three-dimensional Dirac materials satisfies $ln\frac{1}{T}$ within the Hartree-Fock approximation  \cite{Setiawan_Temperature_2015}, which would be suppressed by the presence of strong electron correlation \cite{Das_Many-body_2007}. Here, 
the Fermi velocity $V_{F}$ of the Dirac/Weyl fermions can be obtained from the slope of the interband contribution $\Delta\sigma_{1}$($\omega$), according to 
\begin{equation}
\frac{\Delta\sigma_{1}(\omega)}{\omega} = \frac{NG_{0}}{24V_{F}}  
\end{equation}
 \cite{Hosur_Charge_2012,Tabert_Optical_2016}
where N represents the number of Weyl nodes, and the value of  the quantum conductance $G_{0}$ is 7.74809×$10^{-5}$ s. Since the $\Delta\sigma_{1}$ below 1 015 \cm is contributed by all the Dirac/Weyl cones around the Fermi level, we adopt the slope of this segment to estimate the Fermi velocity, which is almost constant in the whole temperature range. By taking the calculated results of N = 40 \cite{Yang_Transition_2020}, and $\Delta\sigma_{1}/\omega$ = 1.64 $\Omega ^{-1}$, the value of $V_{F}$ is obtained to be 1.48 $\times$ $10^{5}$ m/s. This constant Fermi velocity obviously deviates from the expected $ln\frac{1}{T}$ behavior. However, Similar phenomena have been observed in some weak electron-correlated semimetals, such as ZrTe$_{5}$ \cite{Chen_Optical_2015} and Cd$_{3}$As$_{2}$ \cite{Neubauer_Interband_2016}. Therefore, it is hard to determine the correlation strength of PrAlSi solely by the temperature dependent behavior of the Fermi velocity.

\begin{figure}[ht]
    \centering
    \includegraphics[scale=0.36]{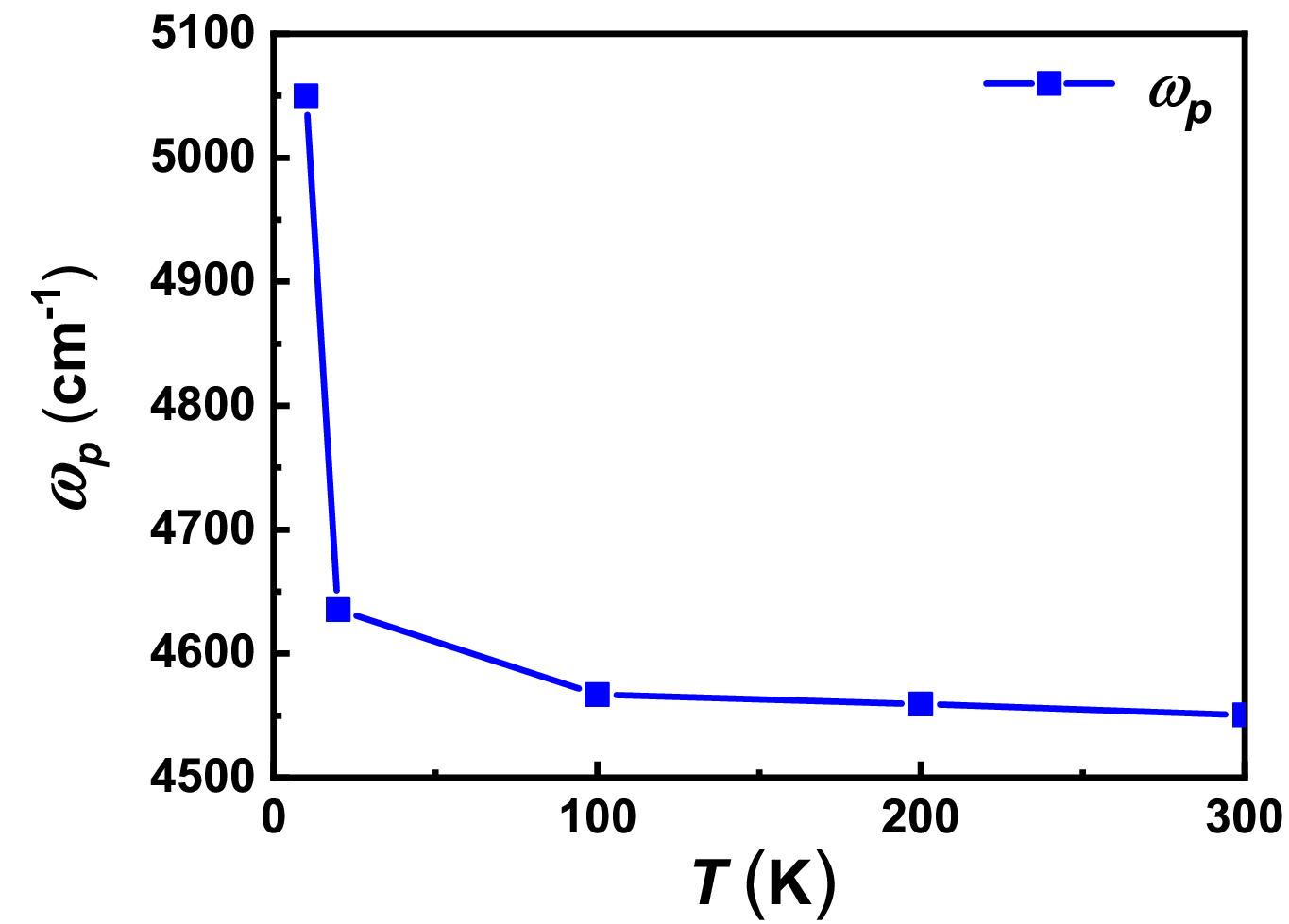}
    \caption{The temperature-dependent plasma frequency $\omega_P$ of PrAlSi.}
    \label{fig4}
\end{figure}

On the other hand, the enhancement of the effective mass $m^{*}$ could be reflected by the plasma frequency $\omega_P$=$\sqrt{4\pi n e^{2}/m^{*}}$. As mentioned above, the value of $\omega_P$ could be obtained by the Drude-Lorentz fitting procedure of $\sigma_1(\omega)$. However, we found that the uncertainties of some parameters are quite large, even though the overall profile is well reproduced. This is likely because the Lorentz model is inadequate in describing the interband transition behaviors of linear dispersive Dirac/Weyl fermions. Therefore, we employ an alternative method to estimate the plasma frequency : 
\begin{equation}
    \omega_{p}^{2} =8\int_{0}^{\omega_{c} } \sigma _{1} (\omega)d\omega,
\end{equation}
where $\omega_{c}$ represents a cutoff frequency that encompasses all intraband transitions while excluding  interband transitions. Here, we designate the positions where $\sigma_{1}(\omega)$ reaches its minimum value between the intra- and interband transitions as the cutoff frequency, which are 602 \cm, 514 \cm, 524 \cm, 603 \cm and 680 \cm  for 10 K, 20 K, 100 K, 200 K and 300 K, respectively.
The obtained $\omega_{p}$ as a function of temperature is displayed in Fig.\ref{fig4}. In consistent with the behavior of $\omega _{p}^{screened}$, the plasma frequency $\omega_{P}$ increase very mildly with decreasing temperature in the normal state, whereas an abrupt enhancement is observed below $T_C$. Furthermore, 
previous Hall effect measurements reveal that the carrier concentrations are of the order of $10^{19}$ $\mathrm{cm^{-3}}$ \cite{lyu_Nonsaturating_2020}. Substituting this value into the equation $\omega_P$=$\sqrt{4\pi n e^{2}/m^{*}}$,
the effective mass is estimated to be $m^{*}\sim 10^{-2}$ $ m_{e}$, where $m_e$ is the free electron mass. Remarkably, it is very rare to observe such small value of $m^{*}$ by infrared spectroscopy, which means the itinerant carriers are almost purely relativistic quasiparticles with an extremely small effective mass. This makes PrAlSi (and probably the whole RSiX family) an exceptional candidate for the development of next-generation devices.  On the other hand, it is difficult to determine whether electron correlation effect has participated in the modification of the extremely small effective mass, which calls for further investigations. A thoroughly understanding of this issue will be crucial for advancing our knowledge of strongly correlated topological materials.

\section{Summary}

In summary, we have studied the charge dynamics of the topological semimetal PrAlSi by infrared spectroscopy. Above the FM transition temperature $T_C$, the interband transition related optical conductivity exhibit two sets of linearly increasing parts, connected by a kink at around 1 960  \cm. We propose they are contributed by neighbouring Dirac/Weyl cones and the kink represents the maximal transition energy between these cones. Below $T_C$, one more linearly growing segment appears in the optical conductivity, giving rise to two kink features at 1 015 \cm and 2 490 \cm. This can be explained by a band structure reconstruction caused by the FM ordering, which transform the Dirac/Weyl cones into pairs of Weyl cones with lower degeneracy. In the meantime, the free carrier density shows a sudden increase across $T_C$, which is probably caused by the shift of Weyl cones relative to the Fermi level during the band structure reconstruction. Additionally, the effective mass of free carriers are identified to be as small as $ 10^{-2}$ $m_e$. These results highlight the strong affection of magnetic orderings towards the topological band structures of PrAlSi, and inspire future theoretical researches on this compound. 

\begin{center}
\small{\textbf{ACKNOWLEDGMENTS}}
\end{center}

We acknowledge the support by the National Key Projects for Research and Development of China (Grant No. 2021YFA1400400 and No. 2022YFA1402200), the National Natural Science Foundation of China (Grant No. 12074042 and No. 12141002), the Young Scientists Fund of the National Natural Science Foundation of China (Grant No. 11704033).  This work was supported by the Synergetic Extreme Condition User Facility(SECUF).


  \bibliography{ReferencesPrAlSi}
\end{document}